\documentclass{aa}
\usepackage{times,graphicx}
\usepackage{txfonts}

\usepackage{color}

\begin{document}


\title{  A unified numerical model of collisional depolarization and broadening rates due to hydrogen atom collisions   }

\author{M. Derouich\inst{1}
   \and A. Radi\inst{2,3}
   \and  P.S. Barklem\inst{4}}

\institute{Sousse University,  ESSTHS, Lamine Abbassi street, 4011 H. Sousse, Tunisia
   \and Ain Shams University, Faculty of Science, Department of Physics, Abbassia, Cairo, Egypt
 \and The British University in Egypt (BUE)  
   \and Theoretical Astrophysics,  Department of Astronomy and Space Physics, Uppsala University, Box 515, S 751 20 Uppsala, Sweden
}

\titlerunning{Collisional  effects}
\authorrunning{Derouich et al.}

\date{Received XXXX / XXXX}

\abstract
{Interpretation of solar polarization spectra accounting for partial or complete  frequency redistribution requires data on various collisional processes.  Data for depolarization and polarization transfer are needed but often missing, while data for collisional broadening are usually more readily available.  Recent work by Sahal-Br\'echot  and Bommier concluded that despite underlying similarities in the physics of collisional broadening and depolarization processes, relationships between them are not possible to derive purely analytically. }
{We aim to derive accurate  numerical  relationships between the collisional broadening rates and the collisional depolarization and polarization transfer rates due to hydrogen atom collisions.  Such relationships would enable accurate and efficient estimation of collisional data  for solar applications.}
{Using earlier results for broadening and depolarization processes based on general (i.e. not specific to a given atom), semi-classical calculations employing interaction potentials from perturbation theory, genetic programming (GP) has been used to fit the available data and generate analytical functions describing the relationships between them.  The predicted relationships from the GP-based model are compared with the original data to estimate the accuracy of the method.}
{Strongly non-linear relationships between the collisional broadening rates and the depolarization and polarization transfer rates are obtained, and are shown to reproduce the original data with accuracy of order 5\%.  Our results allow the determination of the depolarization and polarization transfer rates for hyperfine or fine-structure levels of simple and complex atoms.  }
{In this work we have shown that by using a numerical approach, useful relationships with sufficient accuracy for applications are possible.}


\offprints{M. Derouich, \email{Moncef.Derouich@essths.rnu.tn}}

\maketitle

\section{Introduction}

Interpretation of polarization in solar spectral lines (the so-called second solar spectrum) accounting for   frequency redistribution requires understanding of collisional processes occurring in the solar atmosphere.  In particular, data for depolarization and polarization transfer by collisions with the most abundant perturber, hydrogen atoms, are needed, but often missing and thus neglected in modelling.  On the other hand, data for collisional broadening due to hydrogen collisions are usually more readily available.  Thus, possible relationships between these processes to enable accurate and efficient estimation of depolarization data would be useful in understanding the formation of the second solar spectrum.
 
During the 1990s, Anstee, Barklem and O'Mara (ABO) developed a semi-classical theory for collisional line broadening by neutral hydrogen (Anstee 1992; Anstee \& O'Mara 1991, 1995; Anstee et al. 1997; Barklem 1998; Barklem \& O'Mara 1997; Barklem et al. 1998).  That theory was later extended by Derouich, Sahal-Br\'echot and Barklem (DSB)   to the depolarization and polarization transfer by collisions with neutral hydrogen  (see  for example  Derouich et al.  2003a and Derouich 2004).  The same potentials, based on Rayleigh-Schr\"odinger perturbation theory in the Uns\"old approximation, so called RSU potentials,  were used in both cases.  In addition, the same time-dependent Schr\"odinger equation was solved to obtain the scattering S-matrix; an essential ingredient in the determination of the both collisional depolarization and broadening rates.  

A great advantage of the ABO and DSB  methods is that they are general, i.e. not specific to a given perturbed atom/ion, and therefore adaptable to any neutral or singly ionised atom.  The general, semi-classical methods of ABO and DSB give results in good agreement with accurate but time consuming quantum chemistry calculations
to better than 20\%  for solar temperatures ($T$  $\sim$ 5000 K).  For instance, at $T $=5000K, the difference between results for depolarization rates from the general semi-classical theory (Derouich et al. 2003a) and quantal results   is 11\% for \ion{Mg}{i}  3p $^1P_{1}$, 8\% for \ion{Ca}{i}  4p $^1P_{1}$, and 13\% for Na I 3p $^2P_{3/2}$. 
 In addition, Derouich et al. (2006) showed in their Fig. 6 that the  error bar on their collisional rates determination is well located within the expected error bar
on the value of the solar magnetic field.  They showed that the error bar for the magnetic field determination  due to the typical uncertainty in DSB  
collisional depolarization rates is similar to the error bar for the magnetic field value due to a polarimetric uncertainty of $\sim$ 10$^{-4}$, which corresponds to the uncertainty attainable with polarimeters such as ZIMPOL (Zurich IMaging POLarimeter) and  THEMIS (Heliographic Telescope for the Study of the Magnetism and Instabilities on the Sun).  

The underlying similarity between collisional depolarization and broadening processes as calculated by DSB and ABO suggests the possibility of establishing relationships between them.  The possibility of obtaining such relationships was examined from a theoretical viewpoint by Sahal-Br\'echot  and Bommier (2014).   However, they concluded that there is no possible purely analytical relationship between the collisional depolarization rate and the collisional broadening coefficient, even if the same basic methods and computer codes can be used for obtaining the interaction potentials and collisional scattering matrix.  In this work we aim to provide such relationships between the results of the ABO and DSB methods through a numerical approach.   

The paper is organized as follows. Section 2 gives a brief review of the basic definitions and notations, and describes the collisional data employed in the context of this work. Section 3 explains the optimization approach used to obtain the unified numerical model. The results for simple atoms without hyperfine structure, simple atoms with hyperfine structure and complex atoms  are presented in section 4. Finally, the conclusion of the paper is presented.

 \section{Background and theory}
 
The theory of partial  or complete  redistribution of polarized radiation in a magnetized plasma, which permits to determine the Stokes parameters, must account for the effects of collisions  (e.g. Stenflo, 1994; Bommier, 1997a,b; Landi Degl'Innocenti et al., 1997, Casini et al. 2014). The resultant effect of the collisional depolarization and the collisional broadening by neutral hydrogen is  accompanied by a variety of interesting spectropolarimetry effects.  Collisions act mainly with three distinct, yet correlated, effects: depolarization and polarization transfer,  line broadening, and frequency redistribution in scattering events (partial destruction of the correlation between incident and scattered frequencies).    The theory of collisions and their relationship to the theory of polarization in spectral lines is expounded in detail elsewhere (e.g. Stenflo 1994, Landi Degl'Innocenti \& Landolfi 2004).  Here we simply recall important relations for our work and define our notations.

ABO calculations are performed in the standard dyadic basis (Anstee 1992; Anstee \& O'Mara 1991, 1995; Anstee et al. 1997; Barklem 1998; Barklem \& O'Mara 1997; Barklem et al. 1998).   Let us consider the broadening of the    line profile between
 the initial electronic state  $| n l_1 \rangle$ and the final state $| n l_2 \rangle$ where $n$ is the principal quantum number and $l_1$ and $l_2$ are the   orbital angular
momentum quantum numbers of the states $| n l_1 \rangle$ and  $| n l_2 \rangle$, respectively. We denote by $m_{l_1}$ and $m_{l_2}$  the projections of the orbital angular momentums $l_1$ and $l_2$ on to the interatomic axis, which is taken as the axis of quantization (see ABO papers for more details).  Fig. 1 shows examples of the collisions resulting in the broadening of the line profile for the transition between
 the initial electronic state  $| n l_1 \rangle$ and the final state $| n l_2 \rangle$.  Note, the ABO theory neglects the effects of the spin, and thus fine structure is ignored.  
\begin{figure}[h!]
\begin{center}
\includegraphics[width=6 cm]{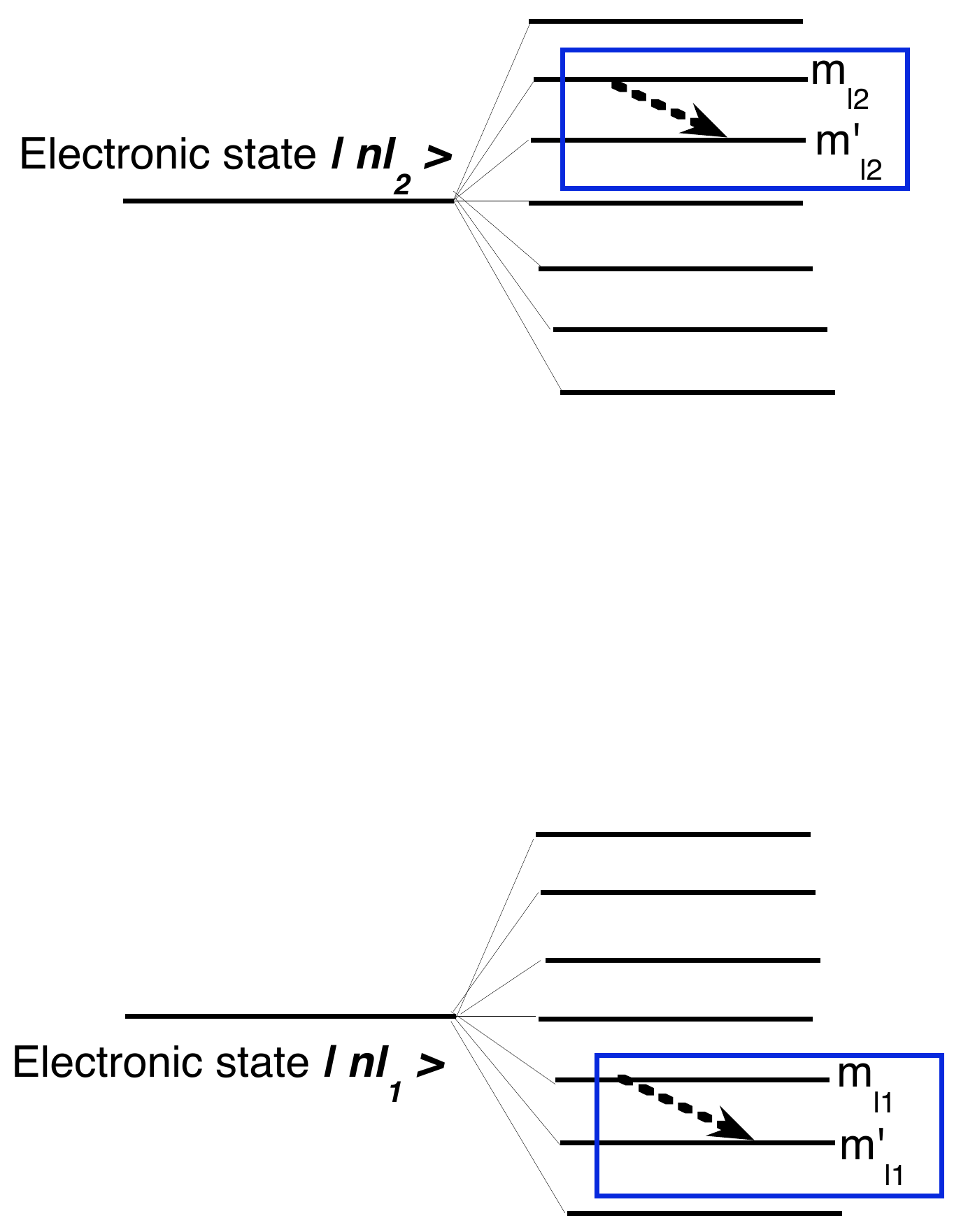}
\end{center}
\caption{As in the ABO theory, we consider   two electronic states $| n l_1 \rangle$ and $| n l_2 \rangle$ connected by a radiative transition. The collisional broadening of the line pofile between 
the levels $| n l_1 \rangle$ and $| n l_2 \rangle$ is a linear combination of all the collisional rates between the Zeeman sublevels inside the state $| n l_1 \rangle$ and the collisional rates between the Zeeman sublevels inside the state $| n l_2 \rangle$. Note that the level spacings are not to scale.}
\label{figure1}
\end{figure}

In order to calculate the depolarization rates   of the fine structure level  $| n l J \rangle$, DSB used the basis of irreducible tensorial operators and  the atomic  states are described by the density matrix elements $ \rho_{q}^{k} (J) $ where the tensorial order $0 \; \le \; k \; \le \;2J$ and the coherence $-k \; \le \; q \; \le \;k$.  More details about the physical   meaning  of the tensorial order $k$ and the coherence $q$ can  be found, for instance, in Omont (1977), Sahal-Br\'echot (1977), and   Landi Degl'Innocenti  \& Landolfi (2004).  We denote by $D^k(J)$   the depolarization rates and by  $\sigma^k_J $   the depolarization cross sections.   The depolarization  rate of the fine structure level  $| n l J \rangle$ is a linear combination of the purely elastic collisions which occur between Zeeman sublevels    $| n l J M_J \rangle$ (see Fig. 2). $M_J$ is  the projection of the total angular momentum  $J$ on the interatomic axis which is taken as the axis of quantization (see DSB papers for more details).  
\begin{figure}[h!]
\begin{center}
\includegraphics[width=6 cm]{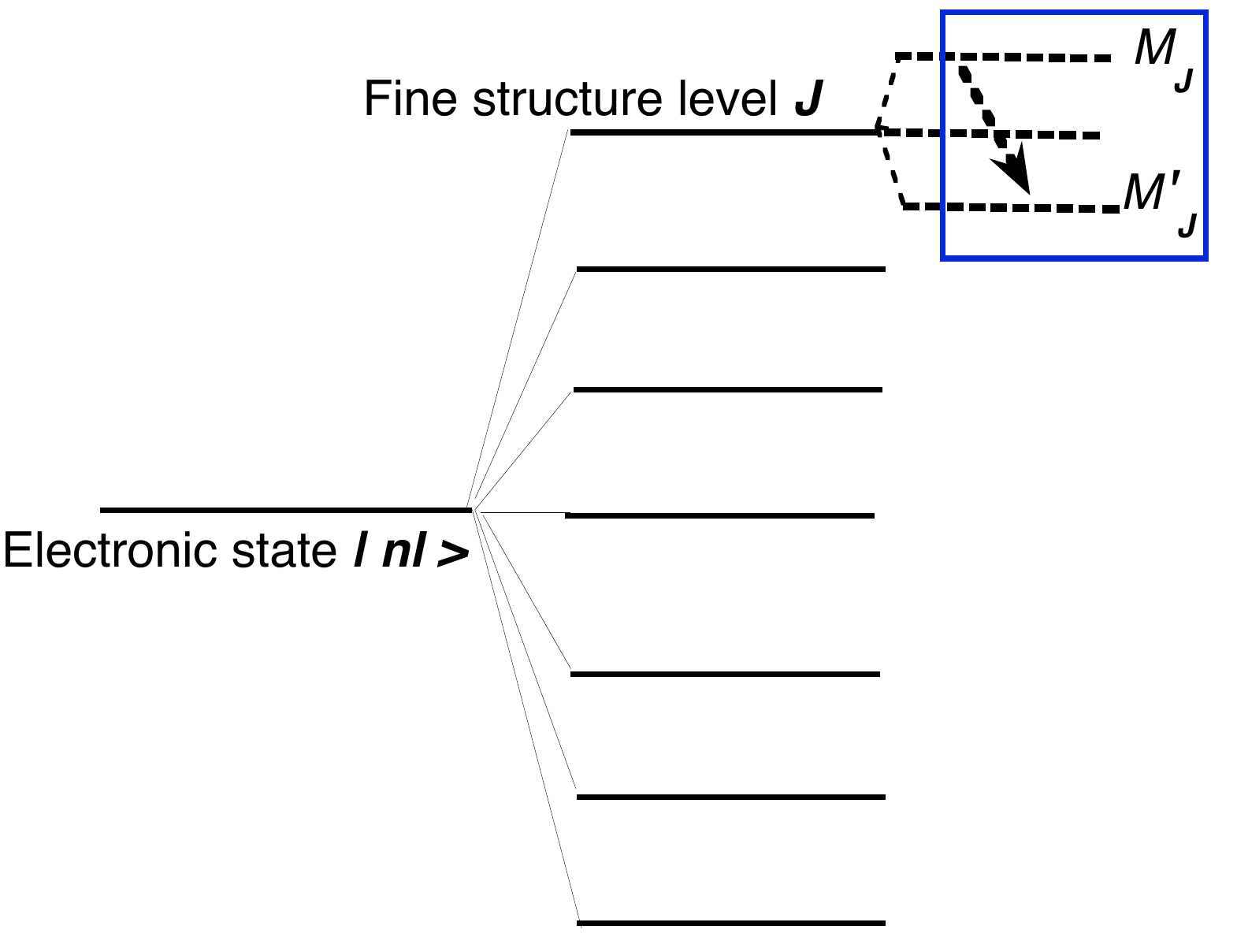}
\end{center}
\caption{ The depolarization rate is a linear combination of the collisional rates between the Zeeman sublevels inside the state $| n l J \rangle$. Note that the level spacings are not to scale.}
\label{figure1}
\end{figure}
 
    The polarization transfer rates from  the $| n l J_1 \rangle$ to the level $| n l J_2  \rangle$    are denoted by  $C^k(J_1 \to  J_2) $;     $\sigma^k_{J_1 \to  J_2} $ are the polarization transfer cross sections. As it is showed in the Fig. 3, transfer  of polarization  between the levels $| n l J_1 \rangle$ and $| n l J_2  \rangle$ is a linear combination 
of the   inelastic collisional rates between the Zeeman sublevels  of the level  $| n l J_1 \rangle$  and  the Zeeman sublevels  of the level  $| n l J_1 \rangle$. Note that, the inealstic collisions which contribute in the  polarization transfer rates occur inside one electronic state. 
\begin{figure}[h!]
\begin{center}
\includegraphics[width=6 cm]{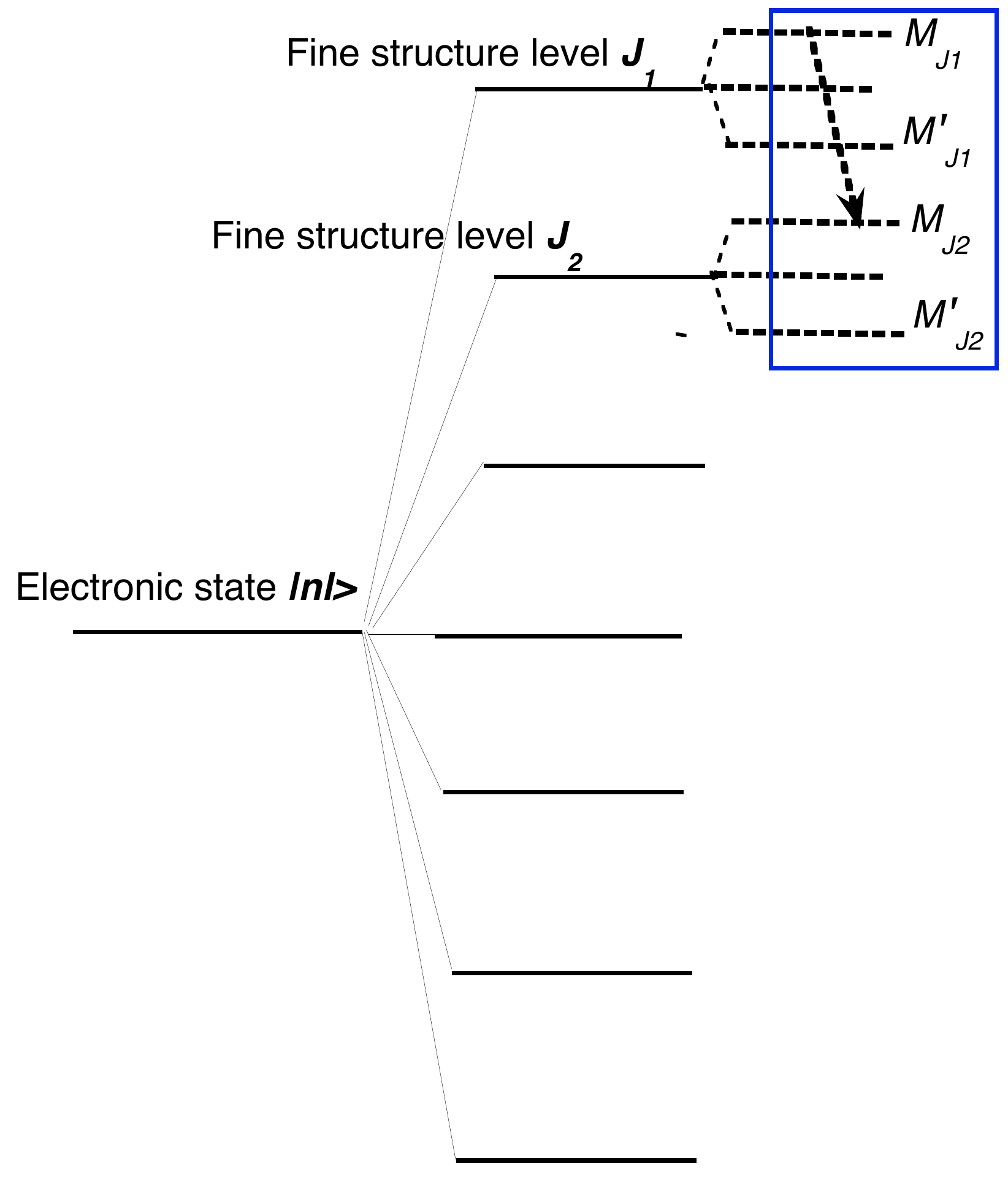}
\end{center}
\caption{ Schematic representation  of   the  collisional transfer  of polarization    between different  fine structure levels.  Note that the level spacings are not to scale.}
\label{figure1}
\end{figure}

The depolarization rates and the polarization transfer rates are :
\begin{eqnarray} \label{eq_1}
D^k(J) =n_{\textrm {\scriptsize H}}   \int_{0}^{\infty}    v \;   f(\vec{v}, T)     \sigma^k_J    (v) , \\
C^k(J \to  J') =n_{\textrm {\scriptsize H}}   \int_{0}^{\infty}    v \;   f(\vec{v}, T)    \sigma^k_{J \to  J'}  (v),
\end{eqnarray}  
 where  DSB found a similar power law relationship with velocity:
 \begin{eqnarray} \label{eq_ch5_4}
 \sigma^k_{J \to  J'}  (v) (J=J'\; \textrm{and}\; J\ne J') &=&
\displaystyle  \sigma^k_{J \to  J'}  (v_0)\left(\frac{v}{v_0}\right)^{-\lambda^k_{J \to  J'}}, 
\end{eqnarray}
with $\lambda^k_{J \to  J'}$ $\sim$ 0.25.  The DSB cross sections, as the ABO cross sections, are tabulated for a reference velocity of $v_0=10^4$~m/s. 

The probability that a collision does not destroy the polarization associated with a $2k$-multipole during a scattering event is (e.g. equation 10.19 of Stenflo 1994):
\begin{eqnarray} \label{eq_1}
p_1  =  \frac{\gamma_E-D^k}{\gamma_N+D^k}, 
\end{eqnarray}  
 where $\gamma_E$ is the elastic collision rate, $\gamma_N$ the radiative broadening rate, and $D^k$ is the rate of depolarizing collisions.  The probability that the   tensorial order  $k$ will be collisionally destroyed
during the scattering process is (equation 10.21 of Stenflo 1994):
  \begin{eqnarray} \label{eq_1}
p_2  =\frac{D^k}{\gamma_N+D^k}.
\end{eqnarray}  
The ratio of these probabilities is
 \begin{eqnarray} \label{eq_1}
R= \frac{p_1}{p_2} =\frac{\gamma_E-D^k}{D^k}  =\frac{1-r}{r}
\end{eqnarray}  
where $r= D^k/\gamma_E$.  Note, $r$ should not be greater than unity since the rate of elastic collisions leading to depolarization may not exceed the  total elastic collision rate.

If one considers an isolated spectral line where inelastic collisions are assumed negligible, and lower state interactions ignored, in the impact approximation the full width at half maximum  (FWHM) of the collisionally broadened line, $\gamma_C$, is equal to the total elastic collision rate $\gamma_E$; see, e.g., pg 517 of Baranger (1962).  In line broadening calculations, including the ABO theory, it is usual to define the broadening cross section such that it gives the half width (HWHM), here denoted $w$ (see below).  Thus, in such a case, $r= D^k/\gamma_E \approx D^k/\gamma_C = D^k/(2w)$, and so in this work we find it convenient to define:
\begin{eqnarray}
R = \frac{2-r_w}{r_w},
\end{eqnarray}
where 
\begin{equation}
r_w =D^k/w = 2r.
\end{equation}

The collisional line width is defined in terms of the broadening cross section $\sigma_w$, by:
\begin{eqnarray} \label{eq_1}
 w = \gamma_c/2 = n_{\textrm {\scriptsize H}}   \int_{0}^{\infty}    v \;   f(\vec{v}, T)   \sigma_w(v), 
\end{eqnarray}  
where $ f(\vec{v}, T) $ is the Maxwellian velocity distribution for a local temperature $T$  and
 $n_{\textrm {\scriptsize H}} $  the perturber  density.  ABO found that  the cross sections have approximately a power law behaviour with velocity, such that they define:
 \begin{eqnarray} \label{eq_ch5_4}
\sigma_w(v)&=& \sigma_w(v_0)\left(\frac{v}{v_0}\right)^{-\lambda}, 
\end{eqnarray}
where $v_0=10^4$~m/s is the reference velocity for which cross sections are tabulated and $\lambda$ is the so-called velocity exponent. ABO found  that  $\lambda$  has a limited range of variation about 0.25 and, as a result, that $w $ has typical resulting temperature dependence of $T^{0.38}$.

We note that several works have attempted to make simple estimates of the relationship between broadening and depolarization.  Smitha et al. (2014) calculated for the case where the interaction between particles can be described by a single tensor operator of given rank, using the equations given by Landi Degl'Innocenti \& Landolfi (2004).  They calculated for their specific case of a \ion{Sc}{ii} line with upper state having $J=2$ for rank 1 and 2 operators, and found that  $r=D^2/\gamma_E=0.1$ and 0.243 (or $r_w=0.2$ and $0.486$), respectively. Interactions of rank 1 correspond to the dipolar interaction, and rank 2 correspond to dipole-dipole interactions; neither is strictly applicable to atom-atom or atom-ion interactions.  Faurobert et al (1995) find $r=D^2/\gamma_C$=0.5 (or $r_w=1$), for the \ion{Sr}{i} resonance line.

The ABO and DSB calculations consider a simple neutral atom having only one optical electron with orbital angular momentum  $0 \le l \le 3$, i.e. $s$-, $p-$, $d-$ and $f$-states.  We note that while the theories can be used for singly ionised atoms, this requires the calculation of the parameter $E_p$ for each state.  For neutrals, this parameter can be approximated by a constant value in all cases, $E_p=-4/9$ atomic units (see, e.g., Anstee 1992, Barklem \& O'Mara 1998).  For this reason, the current work is restricted to neutral atoms.  In the ABO broadening theory, spin is ignored.  However, the calculation of depolarisation requires that spin be accounted for and thus we must consider specific groups of the periodic table. We treat the case of the alkaline earth group where the total angular momentum $J$=$l$. In addition we study the case of the alkali metals where the spin $s$ = 1/2 and $\vec{J}=\vec{l}+\vec{s}$. Therefore, in this paper we consider the cases of the $p$-states where $J=l$=1, $J$=1/2 or 3/2 and $d$-states where $J=l$=2, $J$=3/2 or 5/2 and $f$-states where $J$=$l$=3, $J$=5/2 or 7/2. The case of complex atoms and atoms with hyperfine structure is considered in this work and its treatment is easily derived from the results concerned with simple atoms.  This will be explained in sections~\ref{sect:hyper} and~\ref{sect:complex}.
 
Our goal is to retrieve a numerical relationship between the collisional depolarization rates and the collisional broadening line widths; that is $D^k(J) /w$ and  $C^k(J \to  J')/w$. If one  assumes that the velocity exponents are the same in the case of both broadening and depolarization, i.e. $\lambda^k_{J \to  J'}$ = $\lambda$, which is  approximately the case, one finds that: 
 \begin{eqnarray}  
\frac{D^k(J) }{w}= \frac{ \sigma^k_J    (v_0)}{\sigma_w(v_0)}; \\
\frac{C^k(J \to  J')}{w}=  \frac{ \sigma^k_{J \to  J'}     (v_0)}{\sigma_w(v_0)}.  \nonumber
\end{eqnarray}  
In the next sections we use the tabulated cross sections by ABO and DSB, for a wide range of cases and effective quantum numbers $n^*$, to establish relationships between $\sigma_w(v_0)$, $\sigma^k_J(v_0)$ and $ \sigma^k_{J \to  J'}     (v_0)$.

 \section{Optimization approach}
   To obtain  these  relationships, multipart data analysis and artificial intelligence (AI) techniques are  needed.  AI techniques are becoming very useful as alternate approaches to conventional ones (Whiteson  \& Whiteson 2009).  Within  AI, 
genetic programming (GP) is a global optimization algorithm and an automatic programming technique that  has  been applied in  physics  and astrophysics  (Cohen  et al. 2003,   El-Bakry   \&  Radi 2007, Teodorescu  \& Sherwood 2008, El-dahshan 2009, Schmidt  \& Lipson 2009, Indranil et al. 2013).  GP is a powerful tool that can be used to solve complex fitting problems.  It is a rather recently developed evolutionary computation  method for  deriving  analytical functions  and data analysis.  GP is   based on Darwin's  theory of evolution and uses population of individuals, selects them according to  an accuracy criterion, and produces genetic variation using one or more genetic operators (Koza 1992).   

   GP stores the individuals which are then presented as an expression tree for evaluation  (Oakley  1994).  GP provides the possible solutions to a given optimization problem, using the Darwinian principle of survival of the fittest. It uses biologically inspired operations like replication, recombination and mutation. 
In GP, the chromosome or genome composed of a linear, symbolic string of fixed length, each consists of one or more genes.  Each gene can be represented by an algebraic expression and are composed of a head and a tail. The head contains symbols that represent functions F (with F={sqr,+,-,/,*, or other operators}) and terminals T (with T={x,y,z, or other random constants}), while the tail contains only terminals.  

  As an example, the algebraic expression:  ((x*(x*y))+(x*y)) can be represented as a  tree (Figure 1). Note that    \{+, -, *\} represents the head, and \{x, y, x, z\} represents the tail.  This is the  straightforward reading of the tree from left to right and from top to bottom.  The set of functions and set of terminals must satisfy the closure and sufficiency properties.  The sufficiency property requires that the set of functions and the set of terminals be able to express  the  solution of problem. The function set may contain standard arithmetic operators, mathematical functions, logical operators, and domain-specific functions. The terminal set usually consists of feature variables and constants.  
\begin{figure}[h!]
\begin{center}
\includegraphics[width=6 cm]{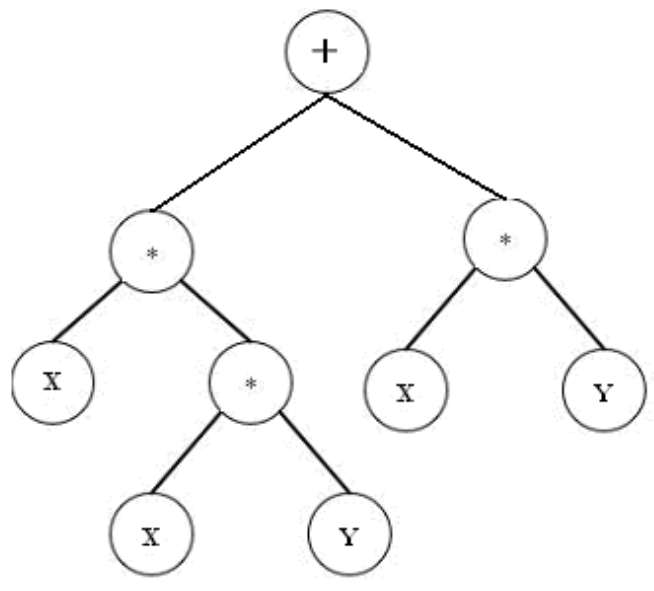}
\end{center}
\caption{Tree representation of the equation +(*(x,y),*(x,*(x,y))) i.e. ((x*(x*y))+(x*y)).}
\label{figure1}
\end{figure}

To measure the error between the actual data found by ABO (denoted by $Z$)  and the data predicted by the numerical model   (denoted by $z_{model}$), we calculated the root of square mean error 
($RSME$) : 
     \begin{eqnarray} \label{eq_opt_2}
        RSME=\sqrt{\frac{\sum\limits_{i=1}^n (Z_{i}-z_{model,i})^2}{n}}
             \end{eqnarray}                                                                            
where $n$ is the number of the values of the data.   The best solution, which is adopted to obtain the results of this paper,  corresponds to the minimum value of $RSME$ (Medhat et al. 2002).
 
 \section{Results and discussion}
 
 ABO presented the broadening cross sections in tabular form for general $s-p$, $p-d$, $d-f$ and $f-d$ transitions. Each cross section was obtained for given  effective principal quantum numbers associated to the upper and lower levels of the transition. For instance, for $p-d$ transitions, broadening cross sections $\sigma_w$ are tabulated, for a relative 
velocity of $10^4$~m s$^{-1}$, with effective principal quantum numbers $n^*_p$ and $n^*_d$.  One can fit  this table  to obtain   an  analytic function of two variables  $\sigma_w$($n^*_p$, $n^*_d$). 
 
DSB tabulated  the variation of the depolarization cross section $\sigma^k$ associated to the $p$, $d$, and  $f$-states, for a relative 
velocity of $10^4$~m s$^{-1}$, with effective principal quantum numbers $n^*_p$,   $n^*_d$, and $n^*_f$. For instance, for the $d$-state, it is possible to obtain  an analytic function of one variable $\sigma^k$($n^*_d$).

The ABO and DSB results are combined in order to obtain analytic functions of two variables:  $\sigma_w$($n^*_s$, $\sigma^k$($n^*_p$)), $\sigma_w$($n^*_p$, $\sigma^k$($n^*_d$)) and $\sigma_w$($n^*_d$, $\sigma^k$($n^*_f$)), for the cases $p$, $d$ and $f$ states, repsectively.   Note that for a transition between a lower level (with effective quantum number $n^*_{l}$) and an upper level  (with effective quantum number $n^*_{u}$), our choice in this paper was to consider $\sigma_w$ as a function of  $n^*_{l}$ and    $n^*_{u}$ and possible depolarization and polarization transfer cross sections $\sigma^k$ associated to $n^*_{u}$. It is also possible to consider $\sigma^k$ associated to $n^*_{l}$, instead of to $n^*_{u}$. However, typically in models it is the broadening associated to ($n^*_{l}$, $n^*_{u}$) and depolarization associated to the upper level $n^*_{u}$, that is of greatest interest. 

In the  following we use GP to treat each possible case and give the corresponding analytic functions.  We note that the most usual situation in astrophysically modelling is that one is able to obtain and estimate for the collisional broadening rate, but requires an estimate for depolarization and polarization transfer rates.  

\subsection{$s$-$p$ and $p$-$s$ transitions}

   For transtions between $s$-states ($l=0$) and $p$-states ($l=1$), the cross sections  $\sigma_w$ are  given in Table 1 of Anstee  \& O'Mara (1991) as functions of the effective quantum number $n^{*}_{s}$ (corresponding to the s-state) and of the   effective quantum number $n^{*}_{p}$ (corresponding to the p-state). For $l=1$, the non-zero  depolarization and polarization transfer cross sections   are $\sigma^2_{J=1} $,  $\sigma^2_{J=3/2} $ and $\sigma^0_{J=1/2 \to 3/2}$.
 These cross sections are given as functions of  $n^{*}_{p}$  in Table 6.1 of  Derouich (2004).  Note that we consider only the tensorial orders $k=0$ and $k=2$ which are relevant to the second solar spectrum studies (i.e. the scattering linear polarization spectrum). 
 We wish to establish the relationships between the function   $\sigma_w$($n^{*}_{s}$,$n^{*}_{p}$) and the  the functions   $\sigma^2_{J=1} (n^{*}_{p}) $,  $\sigma^2_{J=3/2} (n^{*}_{p})$ and $\sigma^2_{J=1/2 \to 3/2} (n^{*}_{p}) $. Since  $\sigma^2_{J=1} (n^{*}_{p}) $,  $\sigma^0_{J=1/2 \to 3/2} (n^{*}_{p})$ and $\sigma^2_{J=1/2 \to 3/2} (n^{*}_{p}) $ depend on $n^{*}_{p}$, the problem becomes to determine the general functions  $\sigma_w$($n^{*}_{s}$,$\sigma^2_{J=1}$), $\sigma_w$($n^{*}_{s}$,$\sigma^0_{J=1/2 \to 3/2}$),  and $\sigma_w$($n^{*}_{s}$,$\sigma^2_{J=1/2 \to 3/2}$).

 \subsubsection{Depolarization rate: $J=J'$=1 and $k=2$ ($s$=0)}
In the case of the depolarization rate   where  the   total angular momentum $J$=$l$=1, the relationship obtained from our GP modeling is:
\begin{eqnarray} \label{eq_depolarization_p}
 Z=-7.+4. \times X/5.+35./Y^2+(245./X) \times (35.-Y) & + & \nonumber \\ (Y \times 75.-X/7.) \times  (Y+5./X) \times (Y-2./Y),
\end{eqnarray}
where $Z$, $Y$, and $X$ are introduced for the sake of clarity  where,   $Z=  \sigma_w(n^{*}_{s},\sigma^2_{J=1})$, $Y=n^{*}_{s}$,   and $X=   \sigma^2_{J=1} (n^{*}_{p}) $. 

To test  the performance of our fitting method we employ ``hare \& hound'' approaches consisting of comparison between the exact solution (called input) and the solution provided by  the equation (\ref{eq_depolarization_p}) (called output).  Figure  2 represents, with a solid line, perfect correspondence (output=input)  and the result of the fit are showed  with squares.  The averaged relative error is $\sim$ 5\%, which is sufficiently precise for applications.
\begin{figure}[h]
\begin{center}
\includegraphics[width=8 cm]{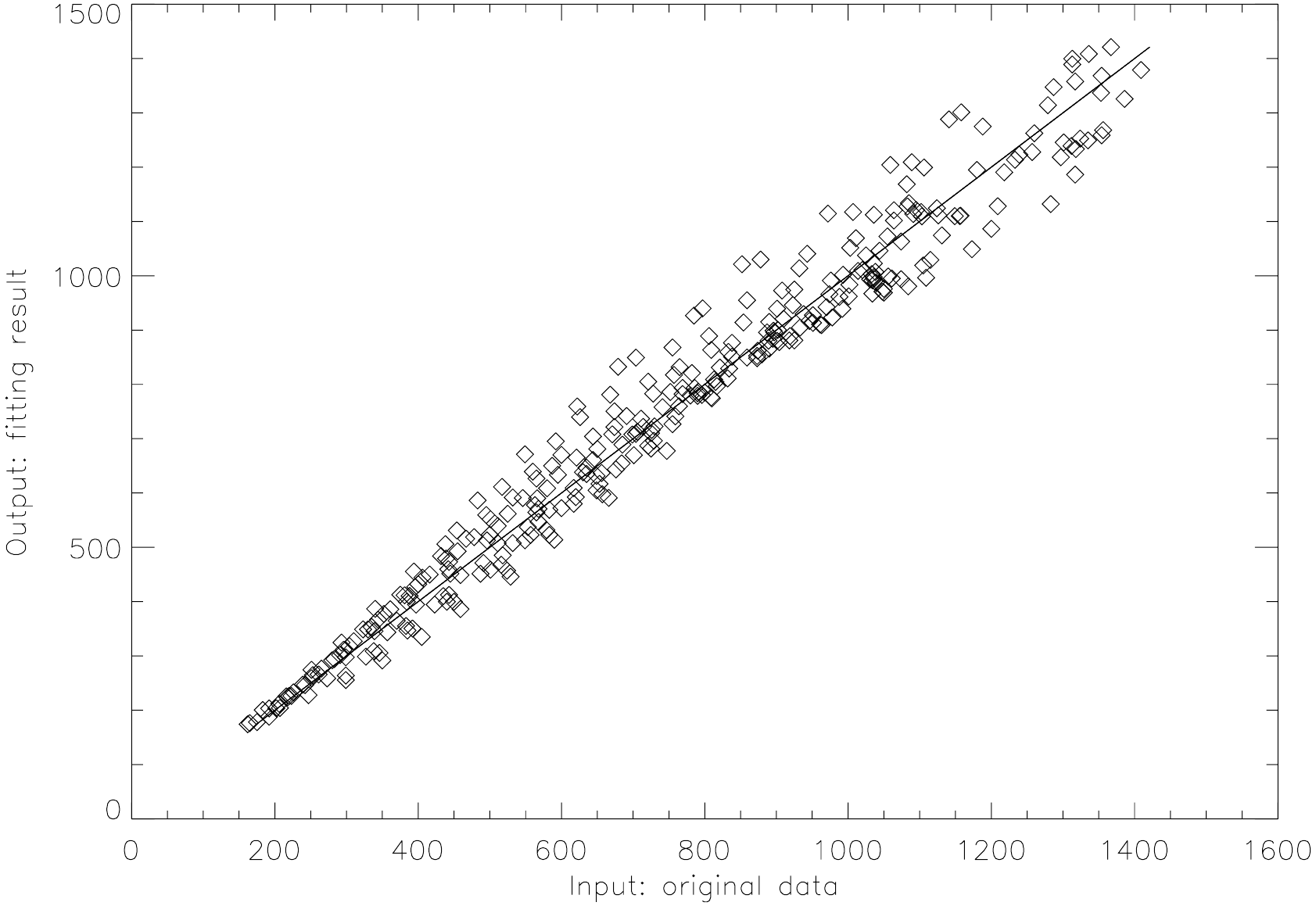}
\end{center}
\caption{To test  the performance of our fitting method we employ ``hare \& hound'' approaches consisting of comparison between the exact solution (called input) and the solution provided by equation (\ref{eq_depolarization_p}) (called output).  The line shows the one-to-one relation.}
\label{figure1}
\end{figure}
 
As an example, we apply Eq. (\ref{eq_depolarization_p}) to determine the relationship between the collisional depolarization and the collisional broadening  rates associated to the photospheric \ion{Sr}{i}  4607 \AA\ line.  The  4607 \AA\ line is the resonance line of \ion{Sr}{i}, namely $5s^2$ $^1S_0$ $\to$ 
$5s$ $5p$ $^1P_1$, and for this line: $n^{*}_{p}(\ion{Sr}{i} )$=2.13 and $n^{*}_{s}$(\ion{Sr}{i} )=1.55.  According to the DSB method (see Derouich 2004 and papers therein), $\sigma^2_{J=1}$=483  a.u. (atomic units). By applying  Eq. (\ref{eq_depolarization_p}), we find that $\sigma_w$(\ion{Sr}{i} )=430 implying that $\sigma^2_{J=1}$=1.12 $\times$ $\sigma_w$ ($r_w=1.12$). We notice that this value is rather similar to the one adopted by Faurobert et al. (1995), where they assumed that   $\sigma^2_{J=1}$= $\sigma_w$ ($r_w=1$) for the \ion{Sr}{i} 4607 \AA\ line.

As mentioned, the more usual case is that one has access to $w$ or $\sigma_w$ and wants to know $D^2$ or $\sigma^2_{J=1}$.  An equation giving X ( $X=   \sigma^2_{J=1}$) as a function of Z (Z=$\sigma_w$) and Y ($Y=n^{*}_{s}$), was therefore determined using our GP model and the  tables of collisional data given by ABO and DSB, namely:
\begin{eqnarray} \label{eq_depolarization_p_inverse}
 X&=&
(2.  \times (Z/7.+1./Y)+Z-Y-6./5)- \nonumber \\  &&  (-2.  \times Y+2  \times Y^2+5.)  \times Y^4
\end{eqnarray}
 In the case of the \ion{Sr}{i} line discussed above, one finds $\sigma_w$(\ion{Sr}{i} )=405~a.u. by interpolation in the tables of Anstee \& O'Mara (1995). Since $Z$=405~a.u. and $Y$=1.55, using the general equation (\ref{eq_depolarization_p_inverse}) one can recover $\sigma^2_{J=1}=480.6$~a.u., a value about 0.5\% lower than the value calculated directly by DSB (see Derouich 2004), 483 a.u. Thus the depolarization and polarization transfer  rates can be obtained via the determination of the broadening  rates. Note that by determining the  depolarization and polarization transfer  rates of simple atoms like \ion{Sr}{i}, the rates associated to complex atoms and atoms with hyperfine structure can be easily obtained (see Sections 4.4 and 4.5 ).  

In the same   homologous  group, one can apply our results to the corresponding transitions in the \ion{Mg}{i}  and \ion{Ca}{i}  atoms. For \ion{Mg}{i}  $3s^2$ and $3p \quad ^1P_{1}$ levels one has  $n^{*}_{s}$(\ion{Mg}{i})=  1.33 and $n^{*}_{p}(\ion{Mg}{i})$=2.03. Then, using Derouich (2004), one has $\sigma^2_{J=1}$ =422  a.u. at $n^{*}_{p}$=2.03. The application of Eq. (10) permits us to determine $\sigma_w$(\ion{Mg}{i} )=361 a.u. and thus $\sigma^2_{J=1}(\ion{Mg}{i})$=1.17 $\times$ $\sigma_w(\ion{Mg}{i})$ ($r_w=1.17$). Similarly, for the case of \ion{Ca}{i}  where  $n^{*}_{p}$(\ion{Ca}{i} )=2.07 and $n^{*}_{s}(\ion{Ca}{i} )$=1.50, one has  $\sigma^2_{J=1}$=444 a.u. and thus  Eq (\ref{eq_depolarization_p}) gives $\sigma_w$=394 a.u.. Therefore, $\sigma^2_{J=1}$=  1.13 $\times$ $\sigma_w$ ($r_w=1.13$).  The results obtained for the resonance lines of \ion{Sr}{i} , \ion{Ca}{i}  and \ion{Mg}{i}, are summarized in  Table 1.
 
\begin{table*}
\begin{tabular}{ccc}
\hline
$r_w$(\ion{Sr}{i} ) & $r_w$(\ion{Ca}{i} ) &  $r_w$(\ion{Mg}{i} ) \\  
\hline
 1.12 & 1.13 & 1.17 \\ 
\hline
\end{tabular}
\caption[]{  $r_w=\frac{D^k(J=1) }{w}= \frac{ \sigma^k_J    (v_0)}{\sigma_w(v_0)}$ for the resonance lines of \ion{Sr}{i} , \ion{Ca}{i} , and \ion{Mg}{i} .} 
\label{table}
\end{table*}
  \subsubsection{Depolarization rate: $J=J'$=3/2, $k=2$ ($s$=1/2)}
   We adopt the same strategy as for $J=J'$=1 in order to determine the ratio $r_w=\frac{D^k(J=3/2) }{w}= \frac{ \sigma^k_{J=3/2}    (v_0)}{\sigma_w(v_0)}$. We obtain the following relationship:
 \begin{eqnarray} \label{eq_Na}
 Z= Y^4  \times  (2.+Y)+49./Y^2+7.  \times  X/(Y+5.)  & + & \nonumber \\
(-588.  \times  Y+98.  \times  Y^3+294.  \times  Y^2)  \times  (5.+Y)  \times  (7./X) 
\end{eqnarray}
 where   $Z=  \sigma_w(n^{*}_{s},\sigma^2_{J=3/2})$, $Y=n^{*}_{s}$,   and $X=   \sigma^2_{J=3/2} (n^{*}_{p}) $.  For the Na I $3  \; s \quad ^2S_{1/2} $ and $3p \quad ^2P_{3/2}$ levels one has $n^{*}_{s}$(Na I)=1.63 and $n^{*}_{p}$(Na I)=2.12. According to Derouich (2004),  $\sigma^2_{J=3/2}$=337 a.u.. Thus, by applying Eq. (\ref{eq_Na}), one finds that $\sigma_w$= 433 a.u. Consequently, $r_w=\frac{D^k(J=3/2) }{w}= \frac{ \sigma^k_{J=3/2}    (v_0)}{\sigma_w(v_0)}=0.777<1.$   In addition, according to  our GP method, the equation giving  $X=   \sigma^2_{J=3/2}$ as a function of $Z=  \sigma_w$ and $Y=n^{*}_{s}$ is: 
 \begin{eqnarray} \label{eq_inverse}
 X=  Z-(Z/Y+5.)/(5. \times Y)-5. \times Y^5
\end{eqnarray}

   \subsubsection{Transfer  of population rate, $J$=1/2, $J'$=3/2, $k=0$}
   The transfer rate of population is denoted by $C^k(J \to  J')$ and the cross section of the transfer   of population is   $\sigma^k_{J \to  J'}$. The relationship between  $\sigma_w$ and $\sigma^k_{J \to  J'} $ is given by:
   \begin{eqnarray} \label{eq_tpopulation}
 Z=(10.-Y^2)/(Y+5.)  \times  ((X/5.+1.-Y)+(X-56.))  \nonumber \\+(1715.  \times  Y^3)  \times  1./X  + 
  (Y+3.)  \times  (Y^3)  \times  (Y+9./4.)
\end{eqnarray}
where,   $Z=  \sigma_w$, $Y=n^{*}_{s}$,   and $X=   \sigma^k_{J \to  J'} (n^{*}_{p}) $. For example, the Na I $3  \; s \quad ^2S_{1/2}$ and $3p \quad ^2P_{3/2}$ levels one has $n^{*}_{s}$(Na I)=1.63 and $n^{*}_{p}$(Na I)=2.12 and using Derouich (2004) one finds that  $\sigma^k_{J \to  J'}$=260 a.u.. The application of Eq. (\ref{eq_tpopulation}) gives  $\sigma_w=433$, thus $r_w=\frac{C^k(J \to  J')}{w}= \frac{ \sigma^k_{J \to  J'} (v_0)}{\sigma_w(v_0)}=0.6$.  Furthermore, by applying  our GP method, the equation giving  $X=   \sigma^2_{J=3/2}$ as a function of $Z=  \sigma_w$ and $Y=n^{*}_{s}$ is: 
 \begin{eqnarray} \label{eq_tpopulation_inverse}
 X=
Z/3.- Y^2 \times Z/14. +3. \times Z/(3./Y^2+5.)
\end{eqnarray}
\subsection{$p$-$d$ and $d$-$p$ transitions}
     \subsubsection{Depolarization rate: $J=J'$=2, $k=2$ ($s$=0)} 
     The depolarization rate for the $d$-states ($l$=2) have been calculated by Derouich (2004) and references therein. In addition the  collisional broadening by neutral hydrogen was obtained by Barklem \& O'Mara (1997). After using our optimization method we obtain:
   \begin{eqnarray} \label{eq_D_depolarization}
 Z=(7. \times X)/(5.+Y)-(1./7.) \times (X/Y)- Y^3 \times 3.5- &&  \nonumber  \\
  490. \times Y^2/((X/5.-49.) \times (5.-Y) \times (3./5.-2./Y))
    \end{eqnarray}
     and 
  \begin{eqnarray} \label{eq_D_depolarization_inverse}
 X &=&
 (Z/7.+Z+2.)/(30./Y)+(Z+4. \times Y)-((2.-Y)/12.)  \nonumber  \\ && \times (Z+5.-Y+5./Z)
     \end{eqnarray}
    where,   $Z=  \sigma_w$, $Y=n^{*}_{p}$,   and $X=   \sigma^2_{J=2} (n^{*}_{d}) $. As an example, let us consider the case $n^{*}_{p}=2$ and $n^{*}_{d}$=3. For $n^{*}_{d}$=3, Derouich (2004) found that $\sigma^2_{J=2}=900$ a.u. Equation (\ref{eq_D_depolarization}) gives $\sigma_w$=820 a.u. The   value given in Table 1 of Barklem \& O'Mara (1997) is $\sigma_w$=834 a.u., the relative error is 1.65 \%. Figure 3 shows the  comparison between the exact solution (the input from Barklem \& O'Mara  1997) with solid line and the solution provided by  the equation (\ref{eq_D_depolarization}) (called output) which  is presented in squares.

\begin{figure}
\begin{center}
\includegraphics[width=8 cm]{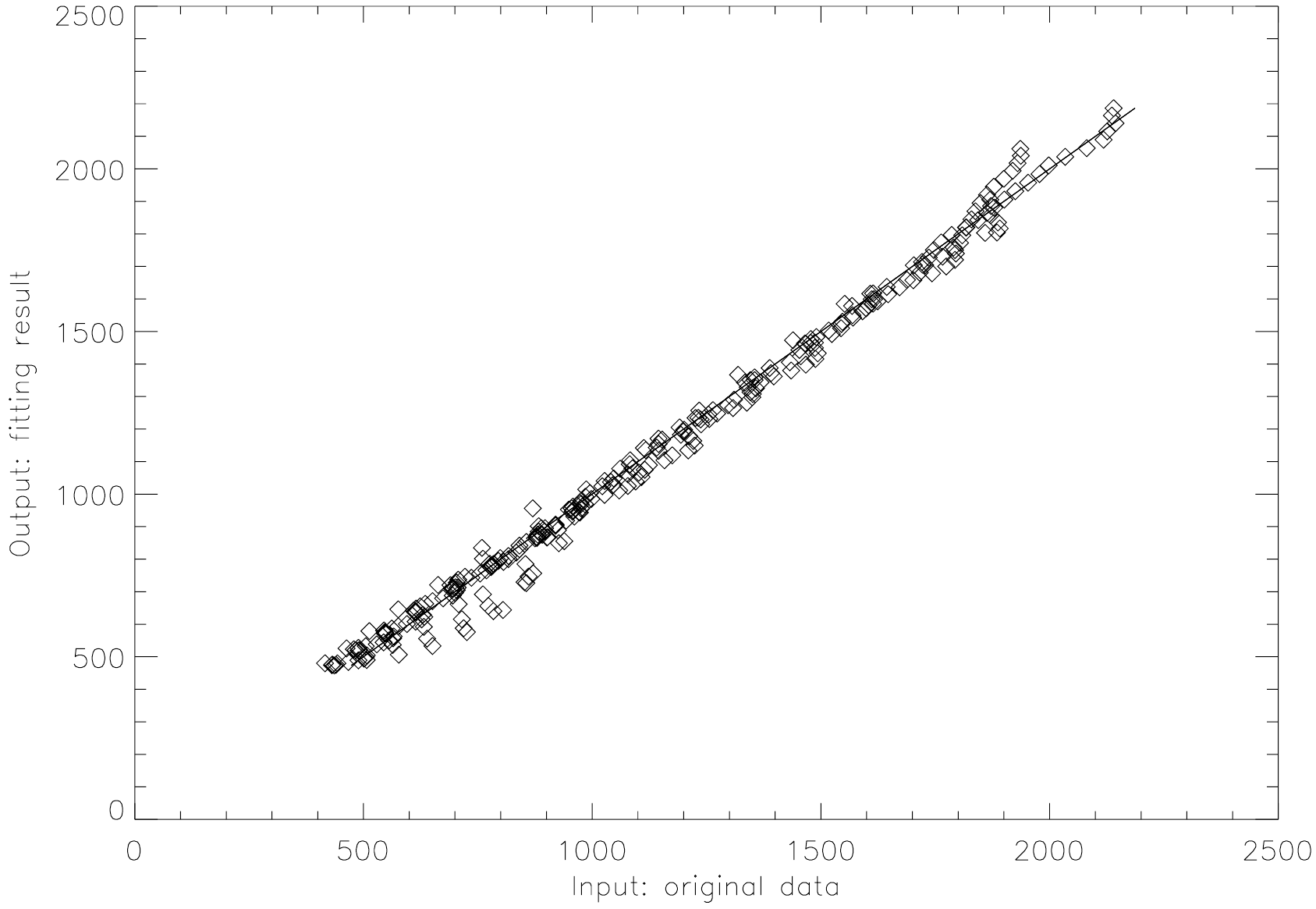}
\end{center}
\caption{To test  the performance of our fitting method we employ ``hare \& hound'' approaches consisting of comparison between the exact solution (called input) and the solution provided by  the equation (\ref{eq_D_depolarization}) (called output). The line shows the one-to-one relation.}
\label{figure1}
\end{figure}
    
One should remark  that the precision of  equations  (\ref{eq_D_depolarization}) and (\ref{eq_D_depolarization_inverse}) are practically the same since the same optimization method and the same data are used to establish these two equations.  This remark is true for all equations of this paper giving $Z$ and $X$. For this reason we usually check the accuracy of the equation giving $Z$. The reader could verify easily that the accuracy of the equation giving X is practically the same as the accuracy of the equation giving $Z$.

     \subsubsection{Depolarization rate: $J=J'$=3/2, $k=2$}   
  The relationship between $\sigma_w$ and $\sigma^2_{J=3/2} (n^{*}_{d}) $ is:
   \begin{eqnarray} \label{eq_D_depolarizationJ32}
Z= X  \times  2./(3.  \times  Y)+247.-49.  \times  Y+X+ &&  \nonumber  \\ (245.  \times  Y  \times  (7.-Y-Y^2))/(-X/7.+Y  \times  6.+16.) 
       \end{eqnarray}
          and 
       \begin{eqnarray} \label{eq_D_depolarizationJ32_inverse}
    X&=& Z  \times Y/(Y+1.)-(35. \times Y) \nonumber  \\ && +  (5. \times (Z-1.)/7.)/(14.+2. \times Y)  \nonumber  \\ && +(Z \times Y/(Y+1.)-3. \times Y)/(8. \times Y+21.)
         \end{eqnarray}
  where,   $Z=  \sigma_w$, $Y=n^{*}_{p}$,   and $X=   \sigma^2_{J=3/2} (n^{*}_{d}) $.  Taking again the case of $n^{*}_{p}=2$ and $n^{*}_{d}$=3, we have    $\sigma^2_{J=3/2}=496$ a.u. which gives  $\sigma_w$=799 a.u.. The relative error in the calculation of the $\sigma_w$ from the relationship of Eq. (\ref{eq_D_depolarizationJ32}) is 4.2 \%.  
      \subsubsection{Depolarization rate: $J=J'$=5/2, $k=2$}   
       Using the optimization method, we find that the relationship between $\sigma_w$ and $\sigma^2_{J=5/2} (n^{*}_{d}) $ is:   
   \begin{eqnarray} \label{eq_D_depolarizationJ52}
Z= X  \times  2./(3.  \times  Y)+247.-49.  \times  Y+X+ &&  \nonumber  \\ (245.  \times  Y  \times  (7.-Y-Y^2))/(-X/7.+Y  \times  6.+16.) 
       \end{eqnarray}
         and 
   \begin{eqnarray} \label{eq_D_depolarizationJ52_inverse}
X &= &  Z+33.-Z/(5.+Y)-Z/(Y+3.)+(Y+7.)  \times  Y
            \end{eqnarray}
      
   where,   $Z=  \sigma_w$, $Y=n^{*}_{p}$,   and $X=   \sigma^2_{J=5/2} (n^{*}_{d}) $.  At $n^{*}_{p}=2$ and $n^{*}_{d}$=3, we have from  Derouich (2004)   $\sigma^2_{J=5/2}=598$ a.u. which gives  $\sigma_w$=   827 a.u.. The relative error in the calculation of the $\sigma_w$ from the relationship of Eq. (\ref{eq_D_depolarizationJ52}) is 0.89 \% and $r_w=\frac{D^k(J)}{w}= \frac{ \sigma^k_{J} (v_0)}{\sigma_w(v_0)}=0.72$.
     
       \subsubsection{Transfer of population rate: $J$=3/2, $J'$=5/2,  $k=0$}      
  We find  the  function giving  the relationship between $\sigma_w$ and $\sigma^2_{J=5/2} (n^{*}_{d}) $:       
   \begin{eqnarray} \label{eq_D_transferk0J32J52}
    Z&=&X/(7.  \times  Y)+(1.-Y)  \times  49.+  X-56.-  (Y-7.)  \times  (X/7.)  \nonumber  \\ && +     (5.  \times  Y^3+Y^5  \times  5.)/[X/(10.   \times   Y)-8.)]
          \end{eqnarray}
         and 
   \begin{eqnarray} \label{eq_D_transferk0J32J52_inverse}
   X&=&
      37.+Z-6./Y-3.  \times  (Z-6.)/(Y+5.)-2  \times Y/3.
       \end{eqnarray}
   where,   $Z=  \sigma_w$, $Y=n^{*}_{p}$,   and $X=   \sigma^0_{3/2 \to 5/2} (n^{*}_{d}) $.   At $n^{*}_{p}=2$ and $n^{*}_{d}$=3, we have from  Derouich (2004)   $\sigma^2_{J=5/2}=512$ a.u. which gives  $\sigma_w$=   821 a.u.. The relative error in the calculation of the $\sigma_w$ from the relationship of Eq. (\ref{eq_D_transferk0J32J52}) is 1.6 \%.  

      \subsubsection{Transfer of alignment rate: $J$=3/2, $J'$=5/2,  $k=2$}      
               
   \begin{eqnarray} \label{eq_D_transferk2J32J52}
    Z=(X^2/(5.+Y))/((1.+Y) \times 6.)+49.   \nonumber  \\   +X/Y+(X+3.) \times 6.+(7./Y^2+24.)/(-18.+X/10.)
          \end{eqnarray}
         and 
   \begin{eqnarray} \label{eq_D_transferk2J32J52_inverse}
    X=
      (Z-3.)/(68./9.+3./Y)-2./Y+10./7.  \times  Y+10.
         \end{eqnarray}
      
   where,   $Z=  \sigma_w$, $Y=n^{*}_{p}$,   and $X=   \sigma^2_{3/2 \to 5/2} (n^{*}_{d}) $.   At $n^{*}_{p}=2$ and $n^{*}_{d}$=3, according to  Derouich (2004),   $ \sigma^2_{3/2 \to 5/2} (n^{*}_{d})=104$ a.u. which gives  $\sigma_w$=    826 a.u.. The relative error in the calculation of the $\sigma_w$ from the relationship of Eq. (\ref{eq_D_transferk2J32J52}) is 1.0 \%. The ratio $r_w=\frac{D^k(J)}{w}= \frac{ \sigma^k_{J} (v_0)}{\sigma_w(v_0)}=0.125$.

\subsection{$d$-$f$ and $f$-$d$ transitions}

     \subsubsection{Depolarization rate: $J=J'$=3, $k=2$ ($s=0$)} 
   The depolarization rate for the $f$-states ($l$=3) have been calculated by Derouich (2004)  and references therein. The  collisional broadening by neutral hydrogen was obtained by Barklem \& O'Mara (1998).
 The analytical relationship between $\sigma_w$ and $\sigma^2_{J=3} (n^{*}_{d}) $ is:       
   \begin{eqnarray} \label{eq_f_depolarization_eq}
   Z=(15.  \times Y^4)/(X+2.) \times (5.  \times Y+35.) \times (Y^2-(7.+Y))  &&    \\
 +(Y-2.)/6. \times (1.+X) \times (3.-Y)+  \nonumber \\
 10./9. \times X+11./9.-Y^2  \nonumber
     \end{eqnarray}
        and 
     \begin{eqnarray} \label{eq_f_depolarization_eq_inverse}
   X &=&
     Z-1./Y+8  \times Y+41.   \nonumber \\ && -(Z-51.-3.  \times Y)/(Y  \times (Y+1./3.))
          \end{eqnarray}
  where,   $Z=  \sigma_w$, $Y=n^{*}_{d}$,   and $X=     \sigma^2_{3} (n^{*}_{f}) $. This equation gives the relationship between  $ \sigma_w$ and $ \sigma^2_{3}  $ for any line, i.e. for all typical values of $n^{*}_{d}$ and $n^{*}_{f}$.  Let us take values of the effective quantum number,   $n^{*}_{d}$=3 and $n^{*}_{f}$=  4. The $\sigma_w$ given by Barklem \& O'Mara (1998) is 1419 a.u.. For the case $n^{*}_{f}$=  4.,  Derouich (2004) found that $\sigma^2_{3}$ ($n^{*}_{f}$=  4)  =1365 a.u.. According to the Eq. (\ref{eq_f_depolarization_eq}) one shows easily  that $\sigma_w=$1464 a.u.. The relative error in the calculation of the cross section via Eq. (\ref{eq_f_depolarization_eq}) is $-$3.1\%.   Figure 4 shows the  comparison between the exact solution (the input from Barklem \& O'Mara  1998) with solid line and the solution provided by  the equation (\ref{eq_f_depolarization_eq}) (called output) which  is presented in squares.

\begin{figure}
\begin{center}
\includegraphics[width=8 cm]{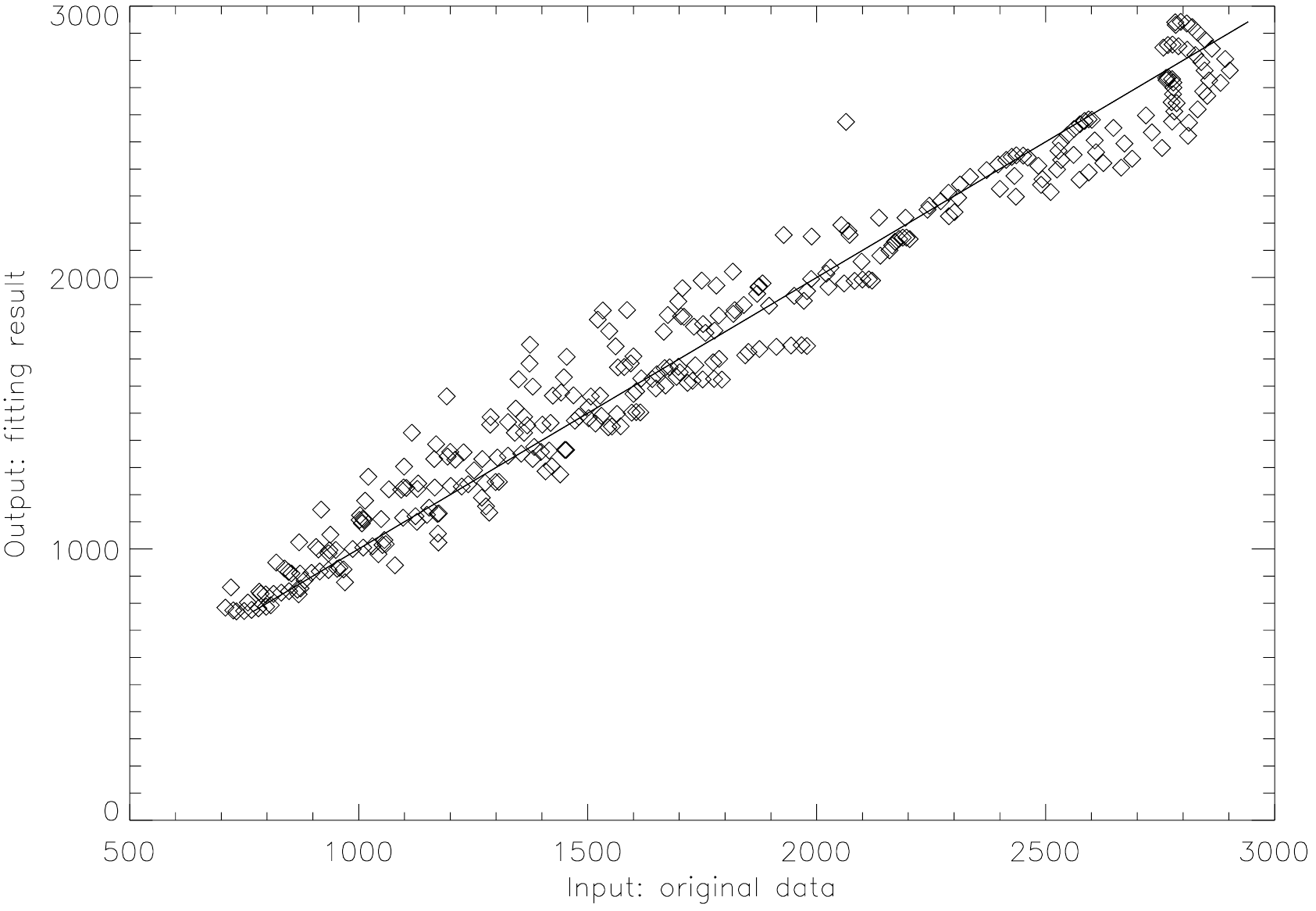}
\end{center}
\caption{To test  the performance of our fitting method we employ ``hare \& hound'' approaches consisting of comparison between the exact solution (called input) and the solution provided by  the equation (\ref{eq_f_depolarization_eq}) (called output). The line shows the one-to-one relation.}
\label{figure1}
\end{figure}
  
 It is worth commenting on the noticeable difference in scatter in Figs. 5, 6 and 7, and the somewhat surprising result that the smallest scatter is seen in Fig. 6, for $p$--$d$ and $d$--$p$ transitions, rather than the scatter say correlating with orbital angular momentum quantum number.  The scatter is largely traceable to non-smooth behaviour in the input data, especially the tabulated broadening cross sections.  Figs. 5, 6 and 7, correspond to the cases of $s$--$p$ (and $p$--$s$), $s$--$p$ (and $p$--$s$), and $d$--$f$ (and $f$--$d$) transitions, respectively, and the behaviour of the cross sections with $n^*$ is displayed graphically in the relevant ABO papers.  There we see that non-smooth behaviour is especially seen at the extremities (i.e. large $n*$) of the tabulated data.  This occurs due to the larger impact of oscillations in the electronic wavefunctions on the calculations.  That is, the wavefunctions have approximately $n^*-l-1$ nodes, and so at around $n* > l+2$ such effects start to be seen in the broadening cross sections.  We note that in the case of $s$--$p$ and $p$--$s$ transitions, the results are tabulated for $s$-states up to $n^*=3=l+3$.  However, in all other cases of the ABO data, the tabulations stop at $n^*=l+2$.   Thus, the rather unexpected behaviour of the scatter is attributed to the larger range of $n^*$ calculated in the $s$--$p$ and $p$--$s$ case.  If we restricted the $s$--$p$ and $p$--$s$ case to $s$-states with $n^*=2$, we could reasonably expect significantly reduced scatter in Fig. 5; see Fig. 1, upper panel, of Anstee \& O'Mara (1995), showing the behaviour is very smooth in this regime.

    \subsubsection{Depolarization rate: $J=J'$=5/2, $k=2$ ($l=3$)}             
The relationship between          $\sigma_w$ and $\sigma^2_{5/2} (n^{*}_{f}) $ is:
  \begin{eqnarray} \label{eq_f_depolarization_J52}
    Z=2. \times X-Y \times 7.-(5.-Y) \times Y \times 5. \nonumber  \\ -(Y-7./Y) \times (Y-2.) \times (X/7.)     \\ +(Y-2.)^2 \times (35./X) \times Y^6  \nonumber
     \end{eqnarray}
        and 
     \begin{eqnarray} \label{eq_f_depolarization_J52_inverse}
    X &=&
     (27./10.+3.  \times Y)  \times (3  \times Y/2.-2.)+   \nonumber \\ && 3./7.  \times Z+(Z+Y)  \times Y/35. 
          \end{eqnarray}
  where,   $Z=  \sigma_w$, $Y=n^{*}_{d}$,   and $X=   \sigma^2_{5/2} (n^{*}_{f}) $=759 a.u..   For $n^{*}_{d}$=3 and $n^{*}_{f}$=  4, we find that  $\sigma_w$=1428 a.u. and that the error in the calculation of the cross section via Eq. (\ref{eq_f_depolarization_J52}) is $-$0.65 \%.  
   \subsubsection{Depolarization rate: $J=J'$=7/2, $k=2$ ($l=3$)}      
       The relationship between          $\sigma_w$ and $\sigma^2_{7/2} (n^{*}_{f}) $ is:
       \begin{eqnarray} \label{eq_f_depolarization_J72}
   Z= X-(10.   \times  Y^2)+(7.   \times  X)/(2.   \times  Y+2.)-(Y+1.)   \nonumber  \\   \times  (7.   \times  X)  \times  (3.-Y)/((5.-Y)   \times  (X/5.-86.)) +2.
     \end{eqnarray} 
 For $n^{*}_{d}$=3 and $n^{*}_{f}$=  4,  the    $\sigma_w$ inferred from Eq. (\ref{eq_f_depolarization_J72})  is 1457  a.u. implying that the percentage of the relative error is $-$2.7 \%. Since $\sigma^k_{J=7/2}$($n^{*}_{f}$=  4)=824 a.u., the ratio $r_w=\frac{D^k(J)}{w}= \frac{ \sigma^k_{J} (v_0)}{\sigma_w(v_0)}=0.565$.          In addition, we find that: 
      \begin{eqnarray} \label{eq_f_depolarization_J72_inverse}
   X &=&  (Z/2.+4.+2 \times Y)-(Z+5.)/(4 \times Y^2)-(-2./Z +2./Z^2)  \nonumber  \\  && \times 5. \times (Z+3.) \times (5.+Y+2. \times Y) 
         \end{eqnarray} 
           \subsubsection{Transfer of population rate: $J$=5/2, $J'$=7/2,  $k=0$ ($l=3$)}  
    We find that:       
            \begin{eqnarray} \label{eq_f_k0_J52_J72}
         Z=2401.   \times  (Y^9)/(X   \times  (X-7.))+(2   \times  X+12.)/(Y+1.) \nonumber  \\  +X  -9.   \times  Y-23.
           \end{eqnarray}    
             and: 
                  \begin{eqnarray} \label{eq_f_k0_J52_J72_inverse}
          X&=& (3.  \times Z/2.+Z  \times Y+8.)/(14  \times Y+82.)+(10./Y)/(Z+1.)   \nonumber \\ &&  +5  \times Y  \times (5.+Y)+(3.+Z)/2.
                      \end{eqnarray}    
where,   $Z=  \sigma_w$, $Y=n^{*}_{d}$,   and $X=   \sigma^0_{5/2 \to 7/2} (n^{*}_{f}) $. For $n^{*}_{d}$=3 and $n^{*}_{f}$=  4, the    $\sigma_w$ inferred from Eq. (\ref{eq_f_k0_J52_J72})  is 1447   a.u.. Therefore, by compring this value to the actual value calculated by Barklem \& O'Mara (1998), one finds   the percentage of the relative error is -2 \%. We notice that, according to DSB, $\sigma^0_{5/2 \to 7/2} (n^{*}_{f}=4.) $=962 a.u.  $<$ $\sigma_w$.       
            
 \subsubsection{Transfer of alignment rate: $J$=5/2, $J'$=7/2,  $k=2$ ($l=3$)}  
  Our final relationship is:
  \begin{eqnarray} \label{eq_f_k2_J52_J72}
         Z=X \times (X-5.)/7.+(7.+X)/(Y-2.)\nonumber  \\+(35.-X/3.) \times ((35.-X/2.) \nonumber  \\ +  
(35.-X/Y) \times (7.-X) \times (2.-Y)/(70.-10. \times Y))
             \end{eqnarray}    
  where,   $Z=  \sigma_w$, $Y=n^{*}_{d}$,   and $X=   \sigma^2_{5/2 \to 7/2} (n^{*}_{f}) $.    For $n^{*}_{d}$=3. and $n^{*}_{f}$=  4., the    $\sigma_w$ inferred from Eq. (\ref{eq_f_k2_J52_J72})  is  1476.6  a.u. implying that the percentage of the relative error is $-4$ \%.  $\sigma^k_{J} $($n^{*}_{f}$=  4)  = 101.    $<$ $\sigma_w$.   In addition, we obtain: 
      \begin{eqnarray} \label{eq_f_k2_J52_J72_inverse}
           X=  7.  \times (8.-Y/Z)+Z/35.-245.245/(Z/21.-30.)
                  \end{eqnarray}    
\subsection{Case of atoms with hyperfine structure}  \label{sect:hyper}

Numerical results presented in the previous subsections are concerned with simple neutral atoms for which the electronic configurations have only one valence electron, and no hyperfine structure. The problem of atoms with hyperfine structure can be easily dealt with by using the fact that, in the framework of the frozen nuclear spin approximation, the hyperfine depolarization and polarization transfer rates are given as a linear combination of the fine rates $D^k(J)$ and $C^k(J \to  J') $   (e.g. Nienhuis 1976 and Omont 1977).  Note that, in typical solar conditions, the frozen nuclear spin approximation is well satisfied.  The hyperfine splitting is usually much lower than the inverse of the typical time duration of a collision and therefore one can assume that the nuclear spin is conserved (frozen) during the collision.

\subsection{Case of complex atoms}  \label{sect:complex}

The electronic configuration of a complex atom  (Fe, Ti, etc.) often has one valence electron outside an open subshell with non-zero angular momentum. We assume that only the valence electron is affected by collisions with hydrogen
atoms (the frozen core approximation). For more details about our approach concerning complex atoms we refer
to Section 2. of Derouich et al. (2005a).  In that paper and in papers that followed (Derouich et al. 2005b, Derouich \& Barklem  2007, Sahal-Br\'echot et al. 2007), it has been shown that the expression for the depolarization rate of a complex atom can be written as a linear combination of the depolarization rates of simple atom. Thus, the relationships obtained and illustrated in this work are applicable to the case of complex atoms provided that one takes into account the formulae giving the depolarization rates of complex atoms as a function of depolarization rates of simple atoms.
   
 \section{Conclusion}
 A coherent  scattering theory which accounts for partial   or complete  frequency redistribution in the presence of isotropic collisions is a challenging problem. Consequently, models of scattering are usually developed in a collisionless regime (e.g. Casini et al. 2014 and references therein).   Stenflo (1994) proposed an approximate model including  redistribution effects and accounts self-consistently for collisions.  Chapter 10 of Stenflo (1994) confronted the problem of including properly the collisional depolarization rates and the collisional broadening coefficients.   

In this paper,  using GP techniques, we obtained accurate relationships for the collisional broadening as functions of the ddepolarization    and polarization tranfer  rates, accurate to typically of order 5\%.    Since the  the more usual case is that one has access to the collisional broadening  and wants to know the depolarization and polarization tranfer rates,  we determined accurate relationships giving the depolarization and polarization tranfer cross-sections as functions of the broadening line width.    As such, these relationships should be useful to the solar polarization community.   

 Interestingly, having the values of broadening coefficients and our relationships it is possible find the depolarization and polarization transfer rates for any hyperfine or fine-structure level of simple and complex atoms.

\end{document}